\documentclass[10pt, conference, compsocconf]{IEEEtran}
\usepackage{cite}
\usepackage{hyperref}

\ifCLASSINFOpdf
\else
\fi

\usepackage{gensymb}
\usepackage{soul}
\usepackage{graphicx}
\usepackage{amsmath}
\usepackage{amssymb}
\usepackage{multirow}
\usepackage{subfigure}
\usepackage[none]{hyphenat}
\usepackage{epsfig}
\usepackage{epstopdf}
\begin{document}
\title{Touchless Typing Using Head Movement-based Gestures}
\author{\IEEEauthorblockN{Shivam Rustagi\IEEEauthorrefmark{1}\IEEEauthorrefmark{2},
Aakash Garg\IEEEauthorrefmark{1}\IEEEauthorrefmark{2}, Pranay Raj Anand\IEEEauthorrefmark{3}, Rajesh Kumar\IEEEauthorrefmark{5}\IEEEauthorrefmark{6}, Yaman Kumar\IEEEauthorrefmark{3}\IEEEauthorrefmark{4} and
Rajiv Ratn Shah\IEEEauthorrefmark{3}}
\IEEEauthorblockA{\IEEEauthorrefmark{2}Delhi Technological University, India \IEEEauthorrefmark{3}Indraprastha Institute of Information Technology Delhi, India\\ \IEEEauthorrefmark{4}Adobe, India \IEEEauthorrefmark{5}Haverford College, USA \IEEEauthorrefmark{6}Syracuse University, USA}
rustagi.shivam3@gmail.com, aakash.garg80@gmail.com, pranay18079@iiitd.ac.in, \\ rkumar@haverford.edu, \{yamank@iiitd.ac.in, ykumar@adobe.com\}, rajivratn@iiitd.ac.in}

\maketitle
\begingroup\renewcommand\thefootnote{\IEEEauthorrefmark{1}}
\footnotetext{These authors contributed equally.}
\endgroup
\maketitle

\begin{figure}[b]
\parbox{\hsize}{\em
The Sixth IEEE International Conference on Multimedia Big Data (BigMM), 2020
\ \copyright 2020 IEEE
}\end{figure}

\begin{abstract}
In this paper, we propose a novel touchless typing interface that makes use of an on-screen QWERTY keyboard and a smartphone camera. The keyboard was divided into nine color-coded clusters. The user moved their head toward clusters, which contained the letters that they wanted to type. A front-facing smartphone camera recorded the head movements. A bidirectional GRU based model which used pre-trained embedding rich in head pose features was employed to translate the recordings into cluster sequences. The model achieved an accuracy of 96.78\% and 86.81\% under intra- and inter-user scenarios, respectively, over a dataset of 2234 video sequences collected from 22 users.
\end{abstract}


% \fi
\IEEEpeerreviewmaketitle

\section{Introduction}
Assistive-technologies have attracted the attention of several Human-Computer Interaction (HCI) researchers over the past decade. People with limited physical abilities find it difficult to interact with traditional equipment, including keyboard, mouse, and joysticks as they require consistent physical-contact based interactions. For example, a person with a physical condition such as Quadriplegia (all four limbs are paralyzed) cannot use any of the off-the-shelf-devices for typing. %\textbf{Before this sentence, you may also add 1-2 lines on other uses/ importance of touchless typing, e.g., games, robotics,  etc. 
%} \hl{Could not find any uses specific to touchless typing in games,roboics that was worth mentioning.}
Recent advances in motion tracking, computer vision, and natural language processing have enabled researchers to develop touchless techniques that assist people in interaction with smart devices. Examples include lip-reading \cite{Salik2019LipperSI}, speech recognition \cite{speechrecog}, Brain-Computer Interface (BCI) \cite{pramitaaai19eeg,64dataset}, eye tracking \cite{eyetype}, and head operated interfaces \cite{thermalspectrum,headwgk}.

The previously proposed touchless technologies suffer from issues such as high error rate, intrusiveness, require expensive equipment, suitable for very specific application scenarios \cite{pramitaaai19eeg,eyetype,64dataset,Salik2019LipperSI,headwgk,thermalspectrum}. For example, speech recognition involves the use of voice cues which are not helpful for a person with speaking disabilities \cite{speechrecog}. Although one of the most convenient methods, the speech recognition system is language-dependent and affected by surrounding noise. Another example includes lip-reading systems which have significant performance discrepancy to speech recognition due to the ambiguous nature of lip actuation, which makes it very challenging to extract useful information \cite{drawbackslipreading}. Similarly, vocabulary and language dependence of lipreading systems is still an active area of research \cite{kumarharnessing}. These problems thus necessitate the use of other types of accessibility devices which overcome these limitations.

BCI-based methods utilize EEG signals generated from a person's brain activity to infer what word/phrase a person was thinking \cite{pramitaaai19eeg}. The capturing of EEG signals is a tedious task as it requires the user to wear a sophisticated device with several electrodes \cite{pramitaaai19eeg,64dataset}. In addition, the BCI devices are expensive, limited to research labs, and are not yet available for common use \cite{pramitaaai19eeg,64dataset}. There are some BCI headbands available in the market such as Muse2\footnote{Muse2 is a multi-sensor meditation device that provides real-time feedback on your brain activity, heart rate, breathing, and body movements to help you build a consistent meditation practice. https://choosemuse.com/muse-2/}, but they contain very limited (four in case of Muse2) electrodes, and are generally used for monitoring sleep. Eye tracking-based typing methods \cite{eyetype} require consistent eye movements which could be strenuous for the user. Head movements-based interfaces have also been explored in the past for controlling a cursor on the screen \cite{head1996} as well as typing \cite{thermalspectrum}. Similar studies utilize face-movement patterns for controlling a cursor, selecting keys, and/or scrolling over rows of the keyboard \cite{facetypewacv}. The majority of these techniques use key selection on modified keyboards. Typing on a modified keyboard poses usability concerns as it could be tedious at times because it generally requires more than one clicks.

\begin{figure*}
 \centering
  \includegraphics[width=7.1in, height =1.2in]{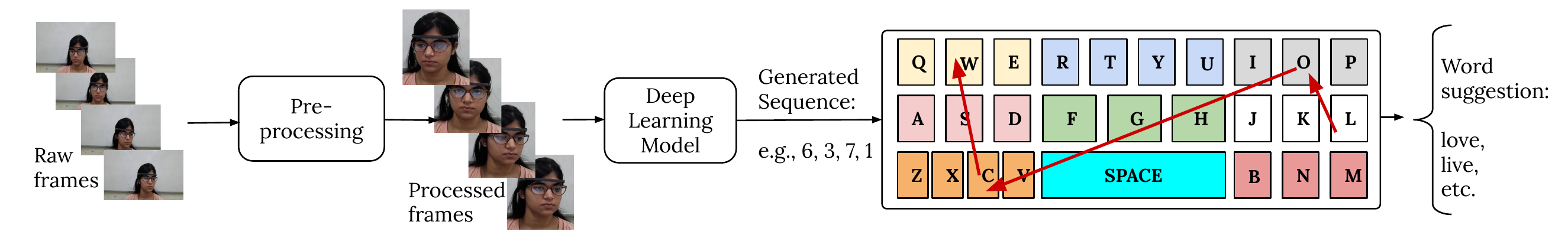}
  \caption{An end to end overview of the proposed touchless typing process. Raw video frames were extracted, preprocessed, and fed to the deep-learning-based generative model, which generated a cluster sequence that was translated to a possible set of words.}
  \label{fig:whole_system}
\end{figure*}

The limitations and drawbacks of previous studies motivated us to explore touchless typing interface that utilizes head movement patterns while the users look at an on-screen QWERTY keypad. Since typing is majorly done using a QWERTY keyboard, it did not take much extra efforts from the users to get accustomed to our proposed touchless typing interface. This also allows for extensibility of the work with different use-cases derived from this. The primary motivation behind our interface choices was usability and cost. Therefore, to capture the head-movement patterns, we used a readily-available and inexpensive smartphone camera (smartphone Samsung M10). Today, this type of camera has both a wide-acceptability and availability.  In addition, similar cameras are almost always available with other mobile devices like laptops, notepads, and smartphones \cite{shah2017multimodal}. The QWERTY keypad was divided into nine color-coded clusters (see the keyboard in Figure \ref{fig:whole_system}) and was displayed on a monitor of 17 inches. In other words, the design of our interface does not require any expensive tracker or a device.

To type a letter sequence, the user makes a series of gestures by looking at the clusters to which the letters belonged to one by one on the virtual QWERTY keyboard. The sequence of gestures captured by the camera is then mapped to a sequence of clusters using deep learning models for sequential data. This predicted sequence of clusters can be used to suggest valid words that can be formed out of this sequence. The process can be compared to touch-enabled swipe keyboard \cite{icaspgoogle} present in modern smartphones. One major difference is the swipe keyboards require the user to touch while our interface does not require the user to touch any interface but to look at virtual QWERTY keypad. Requiring no physical touch make our interface suitable for people with upper limb disabilities. The proposed model in its current form could be used as an assistive tech. However, we believe that the presented interface has the potential to become a mainstream typing technology.

The complete pipeline of our typing system is shown in Figure \ref{fig:whole_system}. Our main contributions are listed as follows:

\begin{itemize}
    \item We develop a dataset of head movement patterns recorded via a central-view camera. The data was collected from a total of 22 users in a lab environment. Each user typed twenty words, ten phrases, and five sentences with average length (number of characters) of 4.33, 10.6 and 18.6, respectively. The dataset and code would be shared publicly for fostering research in this field, further.
    
    \item We propose a bidirectional GRU-based model for translating the recorded videos into a sequence of clusters. The proposed model uses pre-trained embedding rich in head pose features. The performance of the proposed model was evaluated on the aforementioned dataset under inter- and intra-user scenarios using accuracy and Modified Dynamic Time Warping (M-DTW) as performance measures.
    
\end{itemize}

The rest of the paper is organized as follows. Related works, dataset, details of the proposed system and the experimental design is presented in Section \ref{RelatedWork}, Section \ref{Dataset}, and Section \ref{ProposedSystem}, respectively. Section \ref{ResultDiscussion} describes results followed by Section \ref{ConclusionAndFutureWork} that concludes the work beside listing future research directions.

\section{Related Work}  
\label{RelatedWork}
The closely related works can be categorized into selection and gestural-based text entry \cite{vulture}. 

Selection-based text entry mechanisms generally utilize a camera mouse (with some additional mechanism), and an on-screen keyboard \cite{twoletters,CameraMouse,HeadPose}. The additional mechanism includes eye blink, open mouth, etc. and compliments the selection process. In addition, these mechanisms use modified or unconventional keypads. Moreover, they use limited directional (e.g., Left, Right, Up, and Down) movements. As a result, these system exhibit high error rates and are time-consuming. An attempt has been made to reduced time consumption by enabling users to enter a letter in maximum three steps \cite{ThreeStepKeyword}. The individual step consisted of movement of the head in one of the four directions and the return. The text-entry process makes use of a modified keyboard in which the English alphabets were arranged chronologically. The author further improved the system with a two-step entry process \cite{twoletters} and later using a thermal camera to reduce the effect of lighting conditions on the text entry process \cite{thermalspectrum}. 

Word-gesture keyboards (WGK) \cite{wgk} enable faster text entry by allowing users to draw the shape of a word on an input surface. Such keyboards have been used extensively for touch devices. For example, swipe keyboard has been incorporated in the smartphones for a long time and works well. Along the same lines, \cite{vulture} presented a relatively faster mid-air (touchless) typing technique than the selection-based methods. The technique offered comparable performance to gesture-based text entry on touch-enabled input surfaces. The user of the system, however, had to wear a glove with reflective markers that tracked the position of its hands and fingers. To write a word, a user placed the cursor on the first letter, made a pinch gesture using the index finger and the thumb, which followed tracing the remaining letters of the word, and finally releasing the pinch. The authors suggested that the user could select a word from the list of suggestions or continue typing, implicitly confirming that the highlighted word was a match. The user could also undo the suggested word in the text input field by selecting backspace or delete previously typed words by multiple selections of backspace. In other words, they used four basic interaction steps of a match, select, undo, and delete. The authors report that the touchless gesture entry process was 40\% slower than the gesture-based text entry on touch surfaces and mentally demanding. Hand movement-based gestures have also been studied in the context of smartphone security. Researchers have established that hand movements recorded by a front-facing camera can be used to reconstruct smartphone users’ PIN, pattern, and password \cite{HandGestureToPIN,HandGestureToPattern,HandGestureToPassword}.

Besides the aforementioned techniques, one can also use a Virtual Reality Headset (VRH) based method \cite{headwgk} which does not require wearing a pair of hand gloves. The users using the VRH generally control a pointer on a virtual keyboard using head rotation. Another interesting way of touchless typing is through eye-tracking. The eye tracking-based systems detect and track the movement of the pupil to move a cursor \cite{eyepointer} or control a key selector. For example, \cite{9gaze} proposed a system in which they use eye gaze to select keys on a T9 keyboard. Accurate eye gaze systems require sophisticated eye trackers and are also not suitable for long typing sessions as eyes need to be open for a long period.

Our work differs from these works as it uses: a standard QWERTY keyboard, a single mobile camera for recording the head movement-based gestures, and a deep learning-based sequence to sequence model that translate the recorded gestures to a cluster sequence.

\begin{figure}[htp]
    \centering
    \includegraphics[width=3.5in,height = 1.4in]{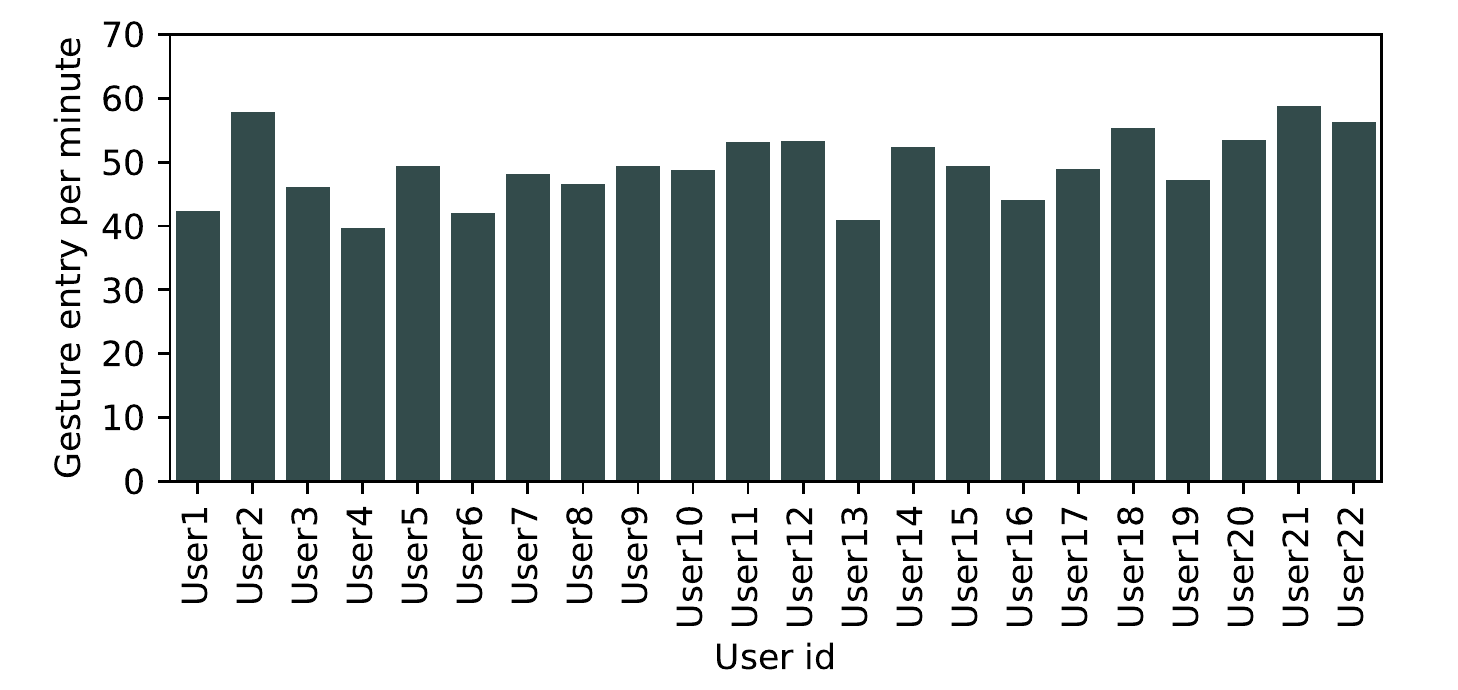}
    \caption{Average number of gesture per minute for each user.}
    \label{fig:gpm}
\end{figure}

\section{Dataset}
\label{Dataset} 
Figure \ref{fig:setup_env} represents the collection setup for multi-modal data consisting of a series of videos and inertial sensor readings. In this section, we describe the dataset setup and collection procedure for the video modality. 
\subsection{Data Description}
\begin{figure}[htp]
    \centering
    \includegraphics[width=2.6in, height = 1.7in]{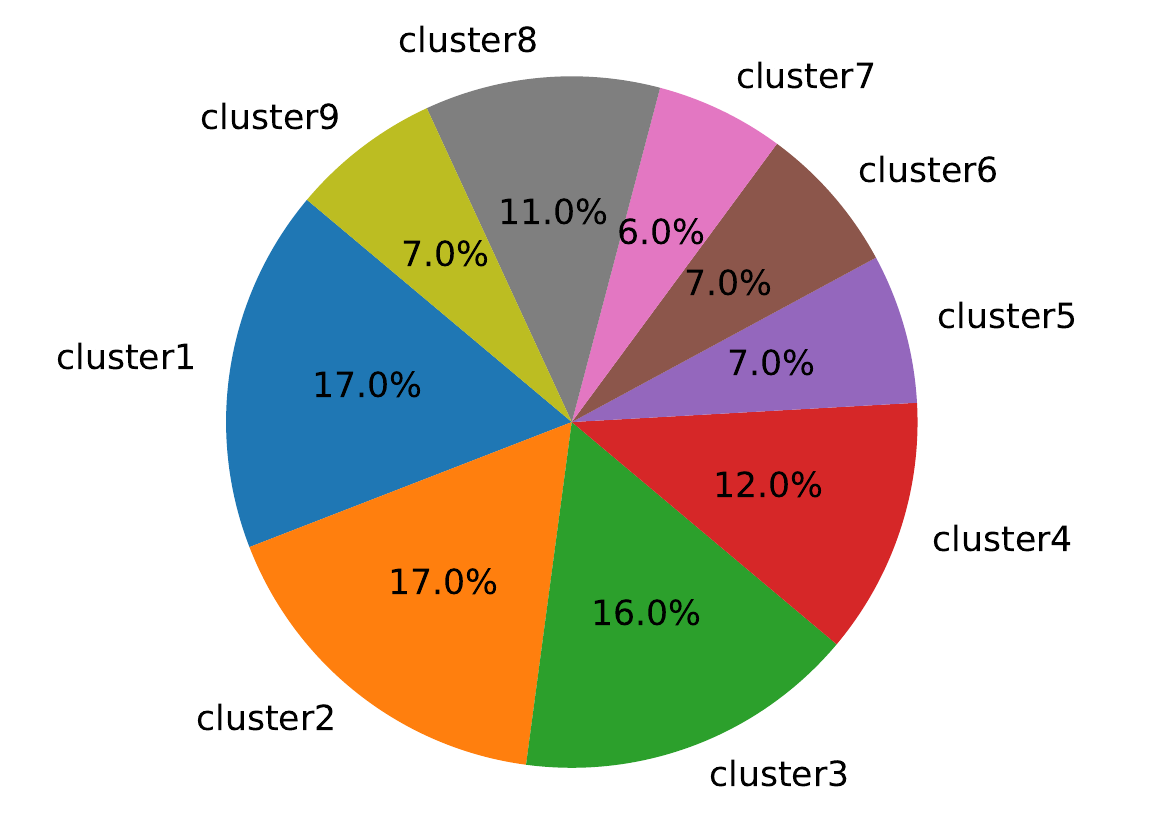}
    \caption{The coverage (share) of each cluster across the dataset. }
    \label{fig:cluster_coverage}
\end{figure}

The dataset consists of 2234 recordings of participants while they typed a number of words, phrases, and sentences by moving their head towards a clustered virtual keyboard displayed on a monitor (see Figure \ref{fig:setup_env}). A total of 25 volunteers participated in the data collection, of which 19 were male, and six were female university students. Out of these, data for three participants were discarded after manual inspection. Each participant typed twenty words, ten phrases, and five sentences (see Table \ref{tab:WordsPhrasesSentences}). The average lengths of words, phrases, and sentences were 4.33, 10.6 and 18.6 letters, respectively. The words were chosen carefully ranging from three to six characters long, such that each of the eight clusters representing 26 English alphabets is included in at least one word, and the unique transitions between the clusters in a given word are maximized (see Figure \ref{fig:cluster_coverage}). The exercise was repeated three times for each participant which resulted in 105 recordings per user. The total number of recordings that were processed were 2310 because some of the samples of two users were not recorded fully and were discarded. The phrases were chosen from OuluVS dataset \cite{ouluvs} and sentences from TIMIT \cite{timit}. The average lengths of words, phrases, and sentences were 4.33, 10.6 and 18.6 letters, respectively. The dataset is a mix of short words like, "ice'', "old'', fly, leg" and large phrases like "hear a voice within you''. Considering every head movement from one cluster to another as a gesture (or a letter entry), the users on an average entered $49.26$ gesture per minute with a standard deviation of $5.3$. The gesture entry rate for each user is show in Figure \ref{fig:gpm} The gesture entry rate would likely increase with more practice on the proposed system. Some examples of mapping from text to clusters are presented in Table \ref{tab:datacollectiondata}.
\begin{table*}[htp]
\caption{List of 20 words, 10 phrases, and 5 sentences that were typed by each participant. The exercise was repeated three times.}
\label{tab:WordsPhrasesSentences}
\centering
\begin{tabular}{|c|c|c|}
\hline
\textbf{Category} & \textbf{Text}                                                                                                                                                               & \textbf{Avg. number of letters per entry} \\ \hline
\textbf{Words}     & \begin{tabular}[c]{@{}c@{}}locate, single, family, would, place, large, work, take, live, \\ box, method, listen, house, learn, come, some, ice, old, fly, leg\end{tabular} & 4.33                                  \\ \hline
\textbf{Phrases}   & \begin{tabular}[c]{@{}c@{}}hello, excuse me, i am sorry, thank you, good bye, see you, \\ nice to meet you, you are welcome, how are you, have a good time\end{tabular}     & 10.6                                  \\ \hline
\textbf{Sentences} & \begin{tabular}[c]{@{}c@{}}i never gave up, best time to live, catch the trade winds, \\ hear a voice within you, he will forget it\end{tabular}                            & 18.6                                  \\ \hline
\end{tabular}
\end{table*}

\subsection{Data Collection Environment}
\begin{figure}[htp]
    \centering
    \includegraphics[width=3.45in, height=2in]{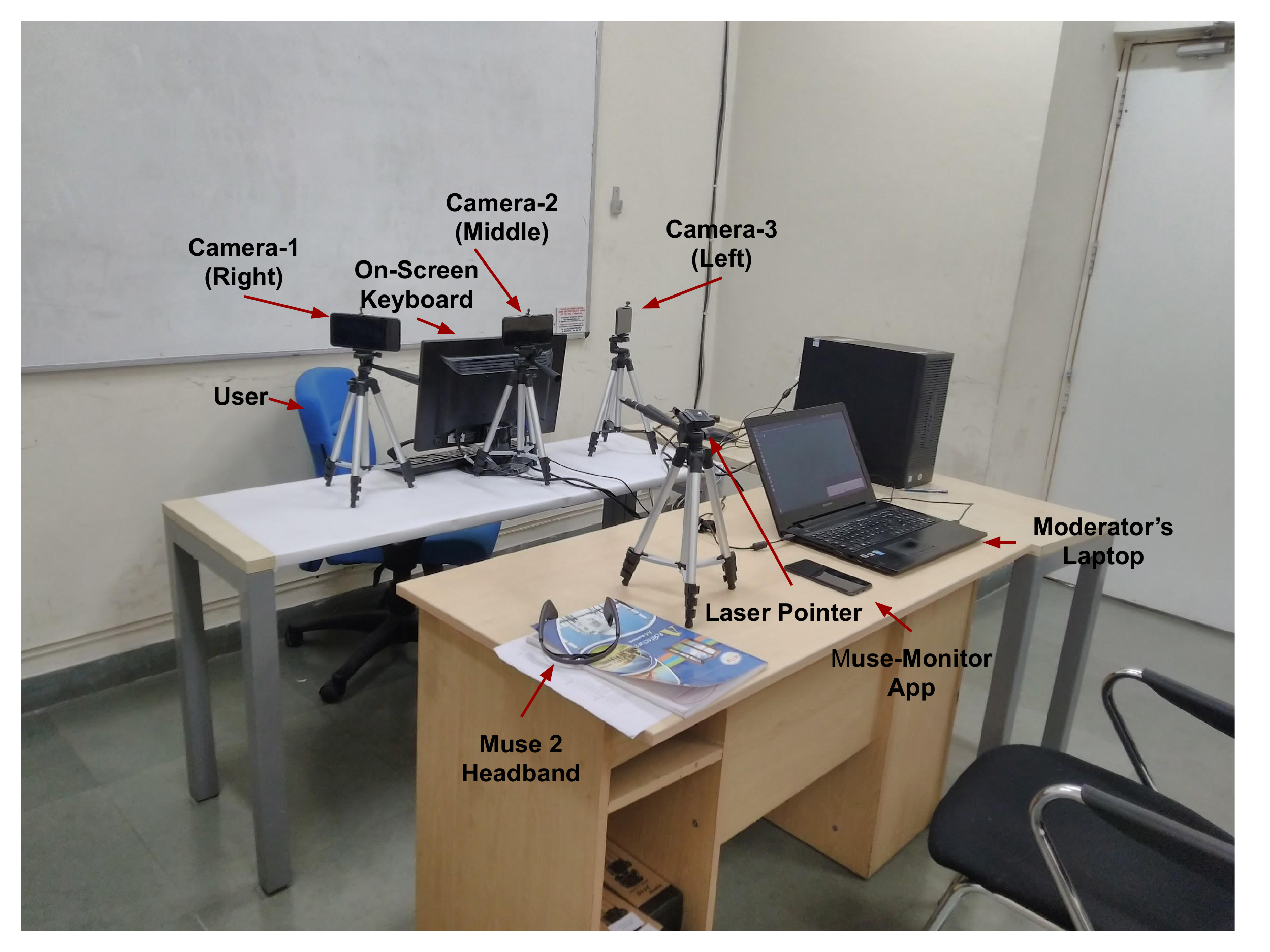}
    \caption{The data collection environment consisting of a monitor on which the virtual keyboard was displayed, three cameras placed on tripods recording the head movement-based gestures of the participant, in addition to Muse 2 which was worn by the participant on his/her forehead, a moderator's laptop, and laser light. All three cameras were facing the participant.}
    \label{fig:setup_env}
\end{figure}

The dataset used in this paper was collected as part of a larger data collection exercise. Figure \ref{fig:setup_env} depicts the overall data collection environment. The goal was to capture the head-movements of the participants via cameras and motion sensors (accelerometer and gyroscope) built into a headband while the users instinctively looked at the cluster one by one (see the keyboard in Figure \ref{fig:whole_system}). Specifically, the data collection setup consisted of three cameras placed on a tripod at -45\degree, 0\degree, and 45\degree, a virtual QWERTY keyboard was displayed on a 17" screen, and a headband (Muse 2) worn by the participant. The camera placed at 0\degree was facing the participant. The three cameras captured the visual aspect of the participant's head movement, whereas headband's accelerometer and gyroscope sensors captured the acceleration and rotation of the head, respectively.

\begin{table}[ht]
\small
\centering
\caption{Examples of word, phrase, sentences, and corresponding cluster sequences.}
\label{tab:datacollectiondata}
\begin{tabular}{|l|l|l|}
\hline
\multicolumn{1}{|c|}{\textbf{Category}} & \multicolumn{1}{c|}{\textbf{Text}} & \multicolumn{1}{c|}{\textbf{Cluster sequence}}   \\ \hline
\textbf{Word}                           & live/love                          & 6, 3, 7, 1                                 \\ \hline
\textbf{Phrase}                         & thank you                          & 2, 5, 4, 9, 6, 8, 2, 3, 2                  \\ \hline
\textbf{Sentence}                       & i never gave up                    & 3,8,9,1,7,1,2,8,5,4,7,1,8,2,3 \\ \hline
\end{tabular}
\end{table}
The recording devices (Samsung M10 smartphones and the Muse 2) were connected and controlled using another laptop (see moderator's laptop in Figure \ref{fig:setup_env}). The videos were recorded at 30 frames per second with a resolution of ${1920 \times 1080}$ pixels. The focus was to record mostly the torso of the participant. A script running on the moderator's laptop broadcast a message to the OpenCamera Remote Apps installed on each of the phones to start and stop the recording simultaneously.

This work utilizes only the information recorded via a central-view camera (the one placed at 0\degree and labelled as Camera2). We believe that the placement of Camera2 (see Figure \ref{fig:setup_env}) setup represents a realistic setup as most of the laptops, notebooks, and phones consist of at least one central-view camera nowadays. On the other hand, the use of multiple cameras and the headband symbolizes a futuristic setup as we envision that the use of multiple cameras and Brain-Computer Interfaces (BCI) headbands would be common for tech-savvy.

The keyboard in Figure \ref{fig:whole_system} demonstrates the color-coded clusters. Each color represents a cluster (a group of nearby keys). The clusters were numbered from 1 to 9. To type a sequence of letters, the participant pointed to the corresponding clusters that consisted of those letters by moving his/her head. For example, if the participant wanted to type the word 'god', the participant would point to the cluster sequence [5, 3, 4]. The use of virtually clustered QWERTY keypad had multiple advantages. First, the participants were already familiar with the keyboard. Second, they did not have to spend much time in locating the keys. They had to point to the clusters by moving their head. Finally, it simplified the problem from predicting 27 keys to predicting nine clusters.

An analysis was conducted to find out how many unique words can be formed from each possible cluster sequence. It was observed that for the 10,000 most common English words\footnote{\url{https://github.com/first20hours/google-10000-english}} there are 8529 unique cluster sequences with each sequence having on an average 1.17 different words. So once we predict the cluster sequence, it can be translated to 1-2 valid words on an average which is comparable to character level prediction. 

The entire data collection exercise was moderated by the research team members who guided the participants as per the need. Specifically, the participants were asked to sit on a chair and rehearse to make themselves familiar with the keyboard and the text entry process. The time of rehearsal varied from participant to participant as they were told to make themselves comfortable with the setup. The moderator reminded the user of the word, phrase, or sentence that was to be typed. The start and stop of the recording was controlled by pressing "Enter" key on the laptop (see Figure \ref{fig:setup_env}). To begin with the participants looked at the centre of the virtual keyboard then started moving their head in the direction of subsequent letters that were to be typed. The recording was stopped once the participant finished moving their head. The process was repeated three times for each of the words, phrases, and sentences.

% \begin{figure}[htp]
%     \centering
%     \includegraphics[width=2in, height=2.6in]{files/actual_setup_pdf.pdf}
%     \caption{A view from the backside of the participant. The participant is wearing a Muse 2 headband and looking at the virtual keyboard displayed on the screen in front of it, and three cameras recording the head-movement gestures. Laser light was used to synchronize the data collection from different cameras as explained in Section \ref{Synchronization}.}
%     \label{fig:setup_diagram}
% \end{figure}

%  following the approval of the Institutional Review Board (IRB) of the university.

\begin{figure*}[t]
 \centering
 \subfigure[HopeNet architecture for predicting yaw, pitch and roll.]{\includegraphics[width=4.6 in, height = 1.8in]{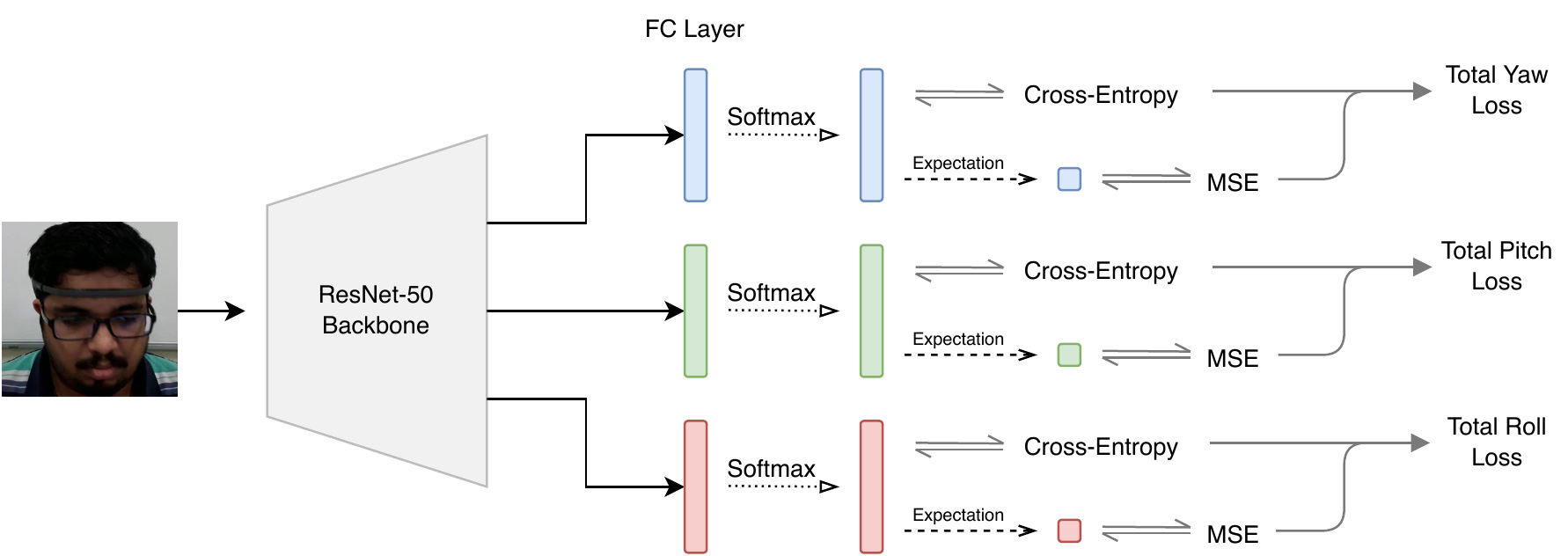}\label{fig:hopenet}} \hfill
 \subfigure[Pose estimation using HopeNet.]{\includegraphics[width=1.7in, height = 1.6 in]{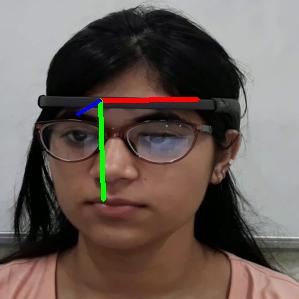}\label{fig:pose_example}}
 \caption{Headpose estimation}
 \label{fig5}
\end{figure*}
%%%%%%%%%%%%%%%%%%%%%%%%%%%%%%%%%%%%%%%%%%
\section{Proposed System}
\label{ProposedSystem}
In this section, we present the entire pipeline of our system, the CNN model used for extracting headpose embeddings, and the GRU based model for sequence translation.\footnote{Code:\url{https://github.com/midas-research/muse-touchless-typing}}

\subsection{System Overview}
The system overview is presented in Figure \ref{fig:whole_system}. First, the region of interest, are extracted from the video using face detection DNN provided in OpenCV \cite{opencv}. Second, the processed video frames are fed into the deep learning models that output the cluster sequence the user was looking at. A set of valid words are recommended based on the predicted cluster sequence.

\subsection{Head Pose Feature Embedding}
For our proposed model, we precompute the features denoting the head pose of a user for each frame separately and train only the RNN model. 
% The problem for human head pose estimation has been categorized into two camps: landmark-based and landmark-free. We didn't use landmark-based methods because these methods depend on the landmark detection, head fitting model, and key points. Secondly, these methods are not robust under scenarios in which the landmark detection fails, in turn, predicts not pose.
HopeNet \cite{ruiz2017finegrained} is a CNN based landmark-free head pose estimation model for computing the intrinsic Euler angles (yaw, pitch, and roll) from an RGB image of a person's face in an unconstrained environment (Figure \ref{fig:pose_example}). The Euler angles are the three degrees used to represent the orientation of a rigid body in 3-dimensional Euclidean space. For predicting the Euler angles, classification, as well as a regression approach, is applied. The angles in the range of  ${\pm}99^{\circ}$ are divided into 66 bins. The network outputs the bin in which the angle lies, and this angle value for the bin is taken as the predicted value for the regression loss. As illustrated in Figure \ref{fig:hopenet}, the method uses a ResNet50 as a backbone network augmented with three fully connected layers of size 66. These three layers use the same ResNet50 backbone with shared weights. HopeNet network uses three loss functions for the three angles. Each one is composed of a coarse-bin classification loss and regression loss. The ground truth feature vector is prepared by the actual Euler values.

For bin classification, softmax cross-entropy loss is used, and for regression loss, the regular mean squared error loss is computed. The final loss function becomes
\begin{equation}\label{eq:hopenet_loss}
    L = H(y, \hat{y}) + \beta \cdot MSE(y, \hat{y})
\end{equation}

Where $L$ represents the total loss, $H$ represents the softmax loss, $MSE$ represents mean squared error loss, $y$ represents the target value, $\hat{y}$ represents the predicted valuer and $\beta$ represents the weight of regression loss.

\begin{figure*}[t!]
    \centering
    \includegraphics[width=7in,height = 2.7in]{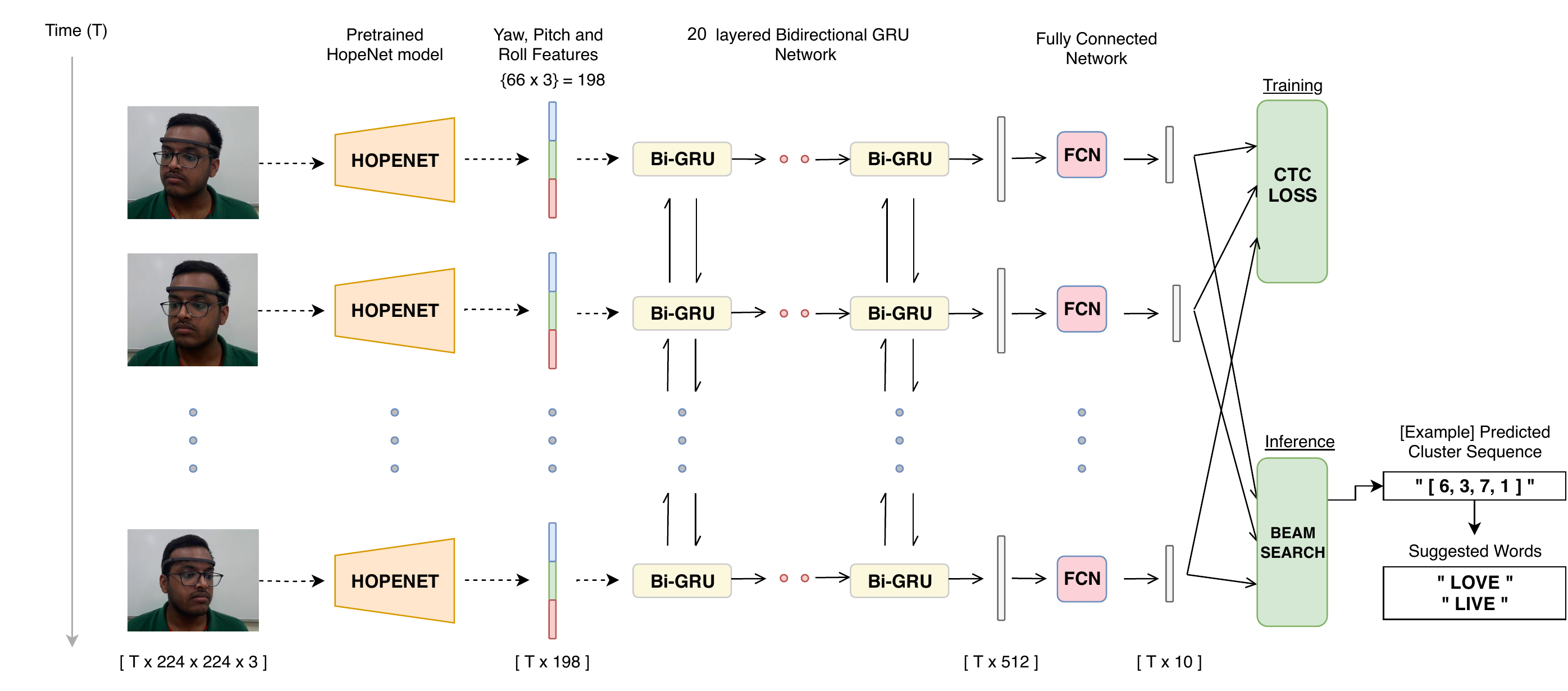}
    \caption{Proposed GRU-based architecture}
    \label{fig:model}
\end{figure*}

Since this model is suitable to form the feature embedding for our task, we use the softmax outputs of size 66 (Figure \ref{fig:hopenet}) for all three Euler angles and concatenate them to form our embedding of dimension 3*66 = 198. 

\subsection{Model Architecture}
The task of generating cluster sequences using head movement video can be seen as a sequence to sequence modelling task in which the input sequence is a sequence of RGB frames and output sequence is a sequence of clusters the user was looking at. Recurrent Neural Networks (RNN) such as LSTM \cite{lstm}, GRU \cite{gru} are widely used for a sequence to sequence modelling tasks. However, it might be the case that the number of video frames is not equal to the number of clusters in the target sequence, and hence an input frame cannot be aligned to a corresponding cluster. For this, we use Connectionist Temporal Classification (CTC) loss \cite{ctc}, which solves the non-alignment problem of input and output. For forming the feature embedding, we use the output from the HopeNet model as described in the previous section.
Our model consists of a twenty-layered Bi-Directional GRU network. The embedding dimension is 198 (concatenation of 66 dimension softmax output for all three Euler angles) and the hidden dimension of the GRU cell used is 512. \\ 
The model architecture is shown in Figure \ref{fig:model}. Features along with the forward and backward directions, each of size 512, are added to give a ${T}\times{512}$ size feature matrix. We apply 1D Batch Normalization followed by a fully connected layer with softmax activation, which reduces the 512 features to ten features. At each timestep, the final output of the model is a softmax vector of size ten consisting of probability at time T for the nine clusters and a blank (for CTC loss). To generate the cluster sequence, we use Beam Search Decoding \cite{wiseman2016sequencetosequence}, a widely used decoding algorithm used in the field of natural language processing.

\section{Training and Performance Evaluation}
\label{TrainingTestingPerformance}
In this section, we describe the training as well as performance evaluation processes.
\subsection{Training}
Instead of training on all N frames of a video, we select every tenth frame, which helps in better capturing the actual directional change during head motion. We use SGD optimizer with nesterov set to true. We use  learning rate = 0.0025 and momentum = 0.9 and set max norm of 400 for tackling gradient explosion. We train the model for 300 epochs on batch size of 20.

\subsection{Performance Evaluation}
The performance of the proposed model was evaluated under two different scenarios which are described as follows:

\subsubsection{Inter-user} In this scenario, the models are trained on a set of users {$S_1$} and tested on a different set of users {$S_2$} such that {$S_1$} and {$S_2$} are mutually exclusive. The cluster sequences are kept the same for train and test sets.

\subsubsection{Intra-user} We had recorded three iterations for each cluster sequence from each user, we trained the models on the first two iterations of each sequence and tested it on the third keeping the users the same in train and test set.

\begin{table}[ht]
\centering
\def\arraystretch{1.5}
%\small
\caption{Performance of the proposed model.}
\label{table:results}
\begin{tabular}{|c|c|c|}
\hline
\multirow{2}{*}{\textbf{Scenarios}} & \multicolumn{2}{c|}{\textbf{Proposed Model}} \\ \cline{2-3} 
                                    & \textbf{Accuracy}      & \textbf{Modified-DTW}      \\ \hline
\textbf{Inter-user}                     & 86.81         & 0.38       \\ \hline
\textbf{Intra-user}                     & 96.78         & 0.07       \\ \hline
\end{tabular}
\end{table}

We evaluated and presented the performance of the model using Accuracy as well as Modified Dynamic Time Warping distance which are described below:

\subsubsection{Accuracy} Accuracy is defined as the ratio of number of correctly decoded cluster sequences and number of total decoded cluster sequences. A cluster sequence is considered correctly decoded if it matches element by element with that of the target sequence. For example, [3, 4, 7, 1, 2] will be considered a correct match to the target sequence [3, 4, 7, 1, 2]. Anything other than [3, 4, 7, 1, 2] will be considered as a mismatch.

\subsubsection{Modified DTW} 
\begin{figure}[htp]
    \centering
    \includegraphics[width=3.2in, height=0.9 in]{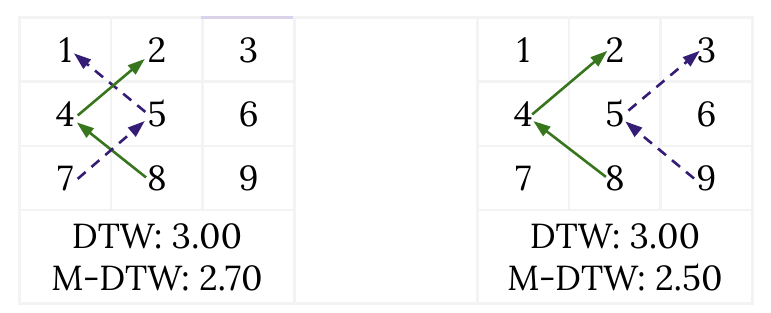}
    \caption{Comparison of DTW and Modified DTW (M-DTW) algorithms. The lesser the distance is, the better the match.}
    \label{fig:DTWComparision}
\end{figure}

Besides Accuracy, we present Modified Dynamic Time Warping (M-DTW) distance \cite{modifieddtwmain} that is specifically designed to compare the sequence of gestures. M-DTW calculates the distance function using a linear combination of Euclidean and Directional distances.
M-DTW was preferred over Standard DTW \cite{dtw} because the standard DTW does not consider the directional changes in 2D space. We consider the keyboard as 2D space and the clusters as coordinates, as shown in Figure \ref{fig:DTWComparision}. Let P = [7, 5, 1], and Q = [9, 5, 3] are two predicted sequences and A = [8, 4, 2] be the actual sequence. The standard DTW distances AP and AQ turned out to be the same. However, in reality, AP should be smaller than AQ as A is more similar to Q than P (see Figure \ref{fig:DTWComparision}). 

\section{Results and Discussion}
\label{ResultDiscussion}
The performance of the model is presented in Table \ref{table:results} under the two experimental scenarios. The model performs exceptionally well when same set of users are used in train and test set and gives an accuracy of 96.78\%. For user invariant scenario, the model gives an accuracy of 86.81\%. The presented results are preliminary as they are based on data collected from a limited number of users under a controlled lab environment. The results are encouraging, and it would be interesting to see how the presented models perform under an unconstrained environment for a much diverse population of users. A factor that would likely improve the performance of the systems is the extent of crisp head-movements by the users during typing. Initial observation suggests that the extent of the head-movements depends on the size of the on-screen keyboard and the distance of the user's head from that screen. 
% \section{Future research directions}
\section{Conclusion and Future Work}
\label{ConclusionAndFutureWork}
This paper presented a touchless typing interface that uses head movement based-gestures. The proposed interface uses a single camera and a QWERTY keypad displayed on a screen. The gestures captured by the camera are mapped to a sequence of clusters using a GRU based deep learning model that consists of pre-trained embedding rich in head pose features. The performance of the interface was evaluated on 2234 video recordings collected from twenty-two users. The presented interface achieved an accuracy of 96.78\%, and 86.81\% under the intra-user and inter-user scenarios, respectively. In the future, the aim is to improve the performance issue by (1) using more training data containing a variety of meaningful sequences, and (2) combining video feeds from multiple cameras, brainwaves recorded via EEG sensors, acceleration, and rotation of the user's head recorded via accelerometer and gyroscope built into Muse 2 which were collected concurrently during the data collection.

We also believe that a new dataset brings with it new challenges. Few of the challenges and opportunities which we observed during the collection and experimentation process are given. Many of the users asked us to have a "click'' functionality which they could potentially indicate by a movement like a long stare or by focussing the eyes on the point of interest. This can also help us to simulate a mouse. Interestingly, this also avoids the time which a mouse takes in moving from one point of the screen to other since motor movements using hand are known to be much slower than eye movements. A significant challenge impacting the usage of the interfaces presented is view and glance dependence. Although we tried to address this by capturing the video from multiple angles, yet we feel more can be done by 3-D modelling of the user viewpoints.

On another note, work in parallel lines can be pursued to define eye movements and head gestures to map them to the various digital movements like right click, left click, crop, copy, paste, \textit{etc.} Other future applications could also work in the direction of integrating the interface with wearable devices and mobile computing. This will bring together a newer set of applications like browsing from wearable glasses.

\bibliographystyle{IEEEtran}
\bibliography{references.bib}


% \end{thebibliography}

\end{document}